\documentclass[journal=jacsat,manuscript=article]{achemso}

\usepackage{siunitx}
\usepackage{color}
\usepackage{graphicx}

\author{Z.\ Yang}
\affiliation{Laboratoire National des Champs Magn\'etiques Intenses,
CNRS-UGA-UPS-INSA, 143, avenue de Rangueil, 31400 Toulouse, France}
\author{A.\ Surrente}
\affiliation{Laboratoire National des Champs Magn\'etiques Intenses,
CNRS-UGA-UPS-INSA, 143, avenue de Rangueil, 31400 Toulouse, France}
\author{K.\ Galkowski}
\affiliation{Laboratoire National des Champs Magn\'etiques Intenses,
CNRS-UGA-UPS-INSA, 143, avenue de Rangueil, 31400 Toulouse,
France}\alsoaffiliation{Institute of Experimental Physics, Faculty
of Physics, University of Warsaw - Pasteura 5, 02-093 Warsaw,
Poland}
\author{A.\ Miyata}
\affiliation{Laboratoire National des Champs Magn\'etiques Intenses,
CNRS-UGA-UPS-INSA, 143, avenue de Rangueil, 31400 Toulouse, France}
\author{O.\ Portugall}
\affiliation{Laboratoire National des Champs Magn\'etiques Intenses,
CNRS-UGA-UPS-INSA, 143, avenue de Rangueil, 31400 Toulouse, France}
\author{R.\ J.\ Sutton}
\affiliation{University of
Oxford, Clarendon Laboratory, Parks Road, Oxford, OX1 3PU, United
Kingdom}
\author{A.\ A.\ Haghighirad}
\affiliation{University of
Oxford, Clarendon Laboratory, Parks Road, Oxford, OX1 3PU, United
Kingdom}
\author{H.\ J.\ Snaith}
\affiliation{University of
Oxford, Clarendon Laboratory, Parks Road, Oxford, OX1 3PU, United
Kingdom}
\author{D.\ K.\ Maude}
 \affiliation{Laboratoire National des Champs Magn\'etiques Intenses,
CNRS-UGA-UPS-INSA, 143, avenue de Rangueil, 31400 Toulouse, France}
\author{P.\ Plochocka}
\email{paulina.plochocka@lncmi.cnrs.fr}\affiliation{Laboratoire National des Champs Magn\'etiques Intenses,
CNRS-UGA-UPS-INSA, 143, avenue de Rangueil, 31400 Toulouse, France}
\author{R.\ J.\ Nicholas}
\email{robin.nicholas@physics.ox.ac.uk}\affiliation{University of
Oxford, Clarendon Laboratory, Parks Road, Oxford, OX1 3PU, United
Kingdom}

\title[An \textsf{achemso} demo]
  {The impact of the halide cage on the electronic properties of fully
inorganic caesium lead halide perovskites}
\abbreviations{IR,NMR,UV}

\begin{document}

\begin{tocentry}

    \includegraphics[width=7.5cm]{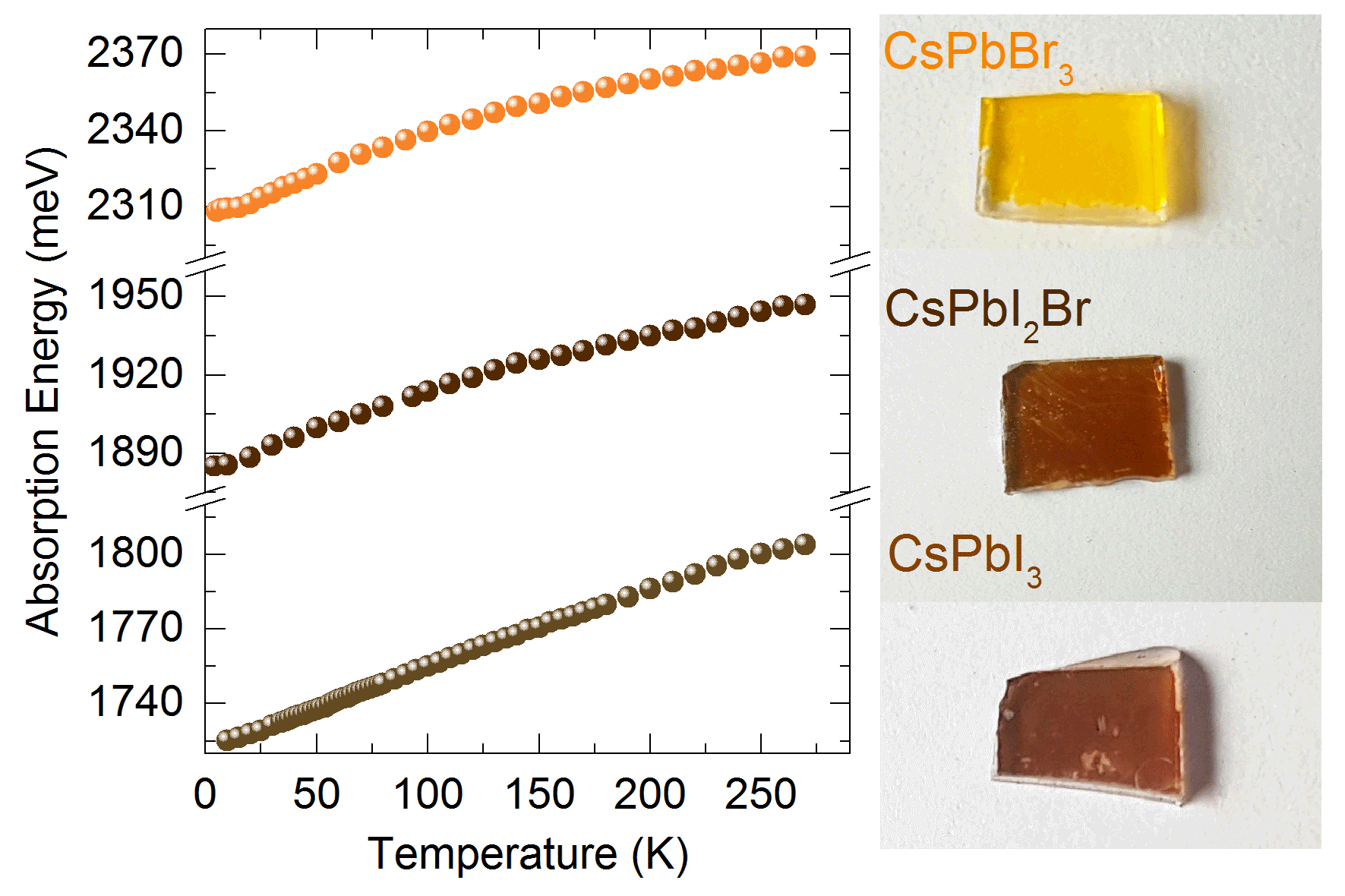}

\end{tocentry}
\newpage
$\textbf{Graphical TOC Entry}$
\begin{figure}[H]
\centering
\includegraphics[width= 7.5 cm]{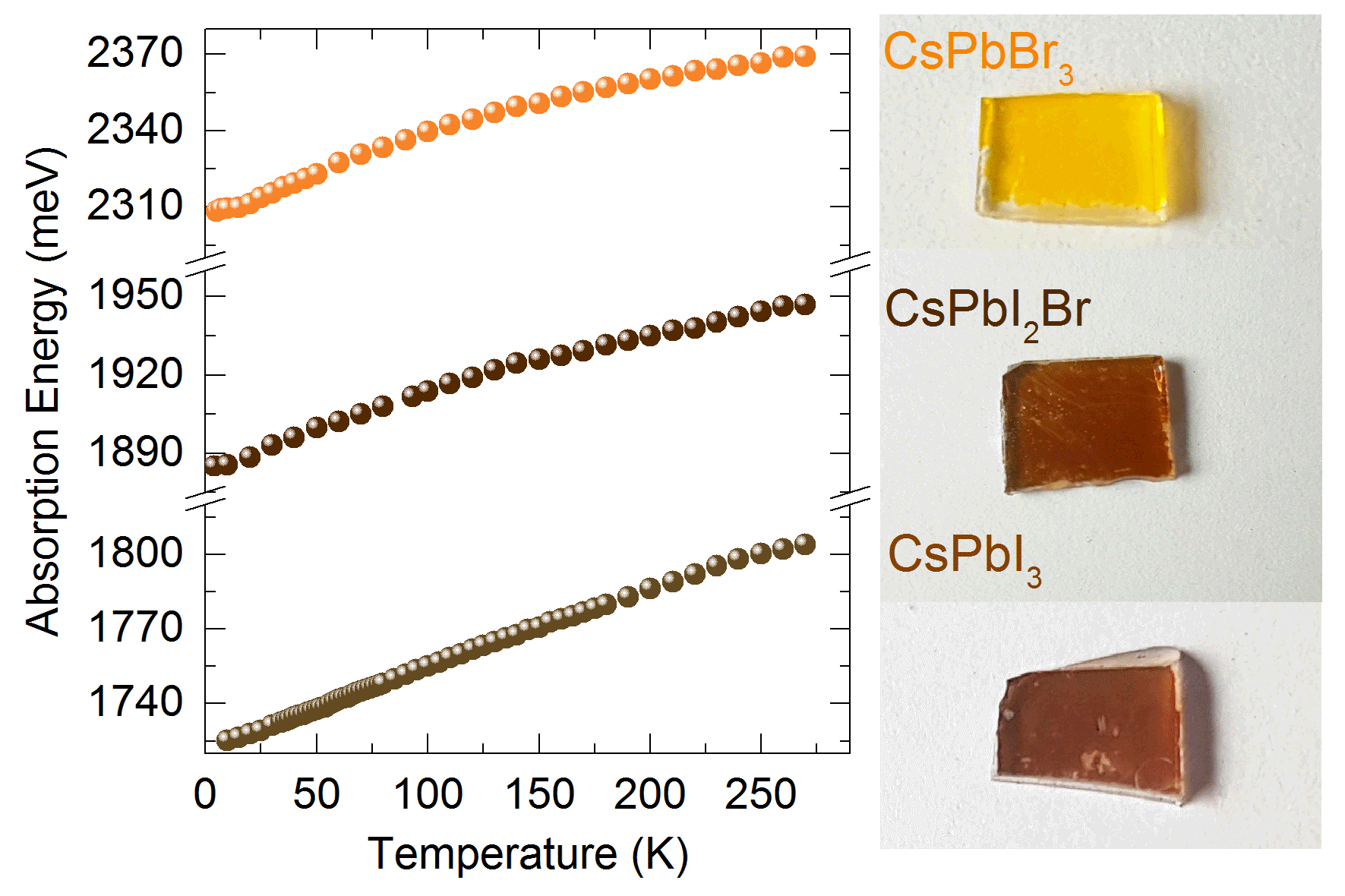}
\end{figure}
\begin{abstract}
Perovskite solar cells with record power conversion efficiency are
fabricated by alloying both hybrid and fully inorganic compounds.
While the basic electronic properties of the hybrid perovskites are
now well understood, key electronic parameters for solar cell
performance, such as the exciton binding energy of fully inorganic
perovskites, are still unknown. By performing magneto transmission
measurements, we determine with high accuracy the exciton binding
energy and reduced mass of fully inorganic CsPbX$_3$ perovskites
(X=I, Br, and an alloy of these). The well behaved (continuous)
evolution of the band gap with temperature in the range $4-270$\,K
suggests that fully inorganic perovskites do not undergo structural
phase transitions like their hybrid counterparts. The experimentally
determined dielectric constants indicate that at low temperature,
when the motion of the organic cation is frozen, the dielectric
screening mechanism is essentially the same both for hybrid and
inorganic perovskites, and is dominated by the relative motion of
atoms within the lead-halide cage.
\end{abstract}

\newpage
Rapid developments in the field of hybrid organic-inorganic
perovskites have led to a dramatic increase of power conversion
efficiencies in perovskite-based solar cells, which currently exceed
22$\%$
\cite{yang2015high,saliba2016cesium,saliba2016incorporation,Shin167}.
Hybrid organic-inorganic perovskites combine low-cost fabrication
processes \cite{burschka2013sequential,liu2013efficient} with strong
light absorption\cite{tanaka2003comparative}, efficient
photoluminescence \cite{deschler2014high,huang2015control}, together
with long carrier lifetimes and diffusion lengths
\cite{stranks2013electron,xing2013long}. The combination of these
properties has led to numerous applications of this class of
materials in optoelectronic devices beyond solar cells, including
light emitting diodes \cite{tan2014bright}, lasers
\cite{zhu2015lead} and photodetectors \cite{fang2015highly}.

Hybrid organic-inorganic perovskites are characterized by a general
chemical formula ABX$_3$, where A is an organic ammonium cation
(Methylammonium (MA), or formamidinium, FA), B$=$Pb$^{2+}$ or
Sn$^{2+}$ and X is a halide anion (Cl$^{-}$, Br$^{-}$, I$^{-}$ or an
alloyed combination of these). Initially, the fabrication of
perovskite-based solar cells was based on mono halide material
\cite{correa2017rapid}.  In this case, the power conversion
efficiency is usually limited to less than 20\% for conventional
MAPbI$_3$ based devices \cite{roldan2015high}, which are
additionally plagued by poor resistance to moisture or high
temperatures \cite{Leijtens2015,misra2015temperature}, as well as by
the formation of trap states induced by exposure to light
\cite{hoke2015reversible}. The synthesis of FAPbI$_3$ provides, in
principle, an attractive alternative, as it has a band gap smaller
than its MA counterpart \cite{koh2013formamidinium}, being closer to
the optimal value for a single junction solar cell, which influences
favorably its conversion efficiency \cite{lee2014high,yang2015high}.
However, the large radius of the FA cation favors the formation of a
photo-inactive polymorph at room temperature
\cite{li2015stabilizing}. Alternatively, fully inorganic
Caesium-based CsPbX$_3$ perovskite compounds \cite{Moller1958} with
an excellent thermal stability up to \SI{450}{\celsius}
\cite{Silva2015,Kulbak2016} have been explored as light harvesters
\cite{Eperon2015,Kulbak2015,Beal2016}. Their use is partially
hindered by the large band gap in the case of CsPbBr$_3$
\cite{Kulbak2015} and by the high temperature formation of the
photoactive polymorph, stable in ambient conditions for the lower
band gap CsPbI$_3$ \cite{Moller1958}.

The advantages of hybrid and inorganic perovskite families can be
combined by introducing some amount of lighter inorganic cations
into the organic lattice. Small molar fractions of Cs act as a
crystallizer for the photoactive FAPbI$_3$ phase
\cite{yi2016entropic,Lee2015}, also reducing halide phase
segregation \cite{McMeekin2016}. The improvement of structural
properties, resulting in superior morphology
\cite{McMeekin2016,saliba2016cesium}, is accompanied by record solar
cell efficiencies and by significantly improved stability under
standard operation conditions
\cite{saliba2016cesium,saliba2016incorporation}. Alloyed Cs- and
FAPbI$_3$ (band gap $\sim\SI{1.73}{\eV}$) has been employed in
perovskite-silicon tandem solar
cells~\cite{Lee2015,Beal2016,McMeekin2016}, while the lower band gap
FA/MA mixtures containing very small molar fractions of Caesium and
Rubidium have demonstrated record high efficiencies exceeding 22$\%$
in a single-junction
architecture\cite{saliba2016cesium,saliba2016incorporation}. Despite
the impressive performance of mixed cation devices, very little is
known about the fundamental electronic properties of these
materials. Moreover, a detailed knowledge of the electronic
structure of fully inorganic perovskites is a crucial element for
understanding the impact of the inorganic cation on the electronic
properties of mixed compounds. Quantities relevant for photovoltaic
applications, such as the exciton binding energy, the presence of
the phase transition and dielectric screening have not as yet been
experimentally investigated in CsPbX$_3$.

In this work, we present the results of systematic studies of the
electronic and optical properties of inorganic CsPbX$_3$ layers,
where X = Br, I or an alloy of the two, which have been heat treated
to be in the metastable cubic black perovskite phase. Temperature
dependent absorption measurements demonstrate no evidence for any
phase transitions below 300\,K in a striking contrast to their
organic counterparts. We determine the the exciton binding energy
($\textit{R}^{*}$) and reduced mass ($\mu$) by magneto transmission
spectroscopy. Similarly to previous results obtained for hybrid
organic inorganic compounds, we find that both exciton binding
energy and reduced mass scale linearly with the band gap energy
\cite{Miyata15,Galkowski2016}. The small binding energy of the
exciton in CsPbI$_3$ compared to the thermal energy at room
temperature suggests that in typical operational conditions for
solar cells, photo-created species exhibit a free-carrier-like
behavior, which makes CsPbI$_3$ an excellent building block for
stable, high efficiency devices
\cite{saliba2016cesium,saliba2016incorporation}. Based on the value
of binding energy, we have calculated the dielectric constant
($\varepsilon_{\text{eff}}$) and compared it with the values for
organic inorganic counterparts. Interestingly,
$\varepsilon_{\text{eff}}$ is comparable for all the iodide
compounds, but decreases significantly for the bromides. This
suggests that at low temperature, when the motion of the organic
cations is frozen \cite{Poglitsch87}, the dielectric screening
mechanism is essentially the same for both the inorganic and hybrid
perovskites and is controlled by the lead-halide cage.

\begin{figure}[]
\centering
\includegraphics[width= 1.0\linewidth]{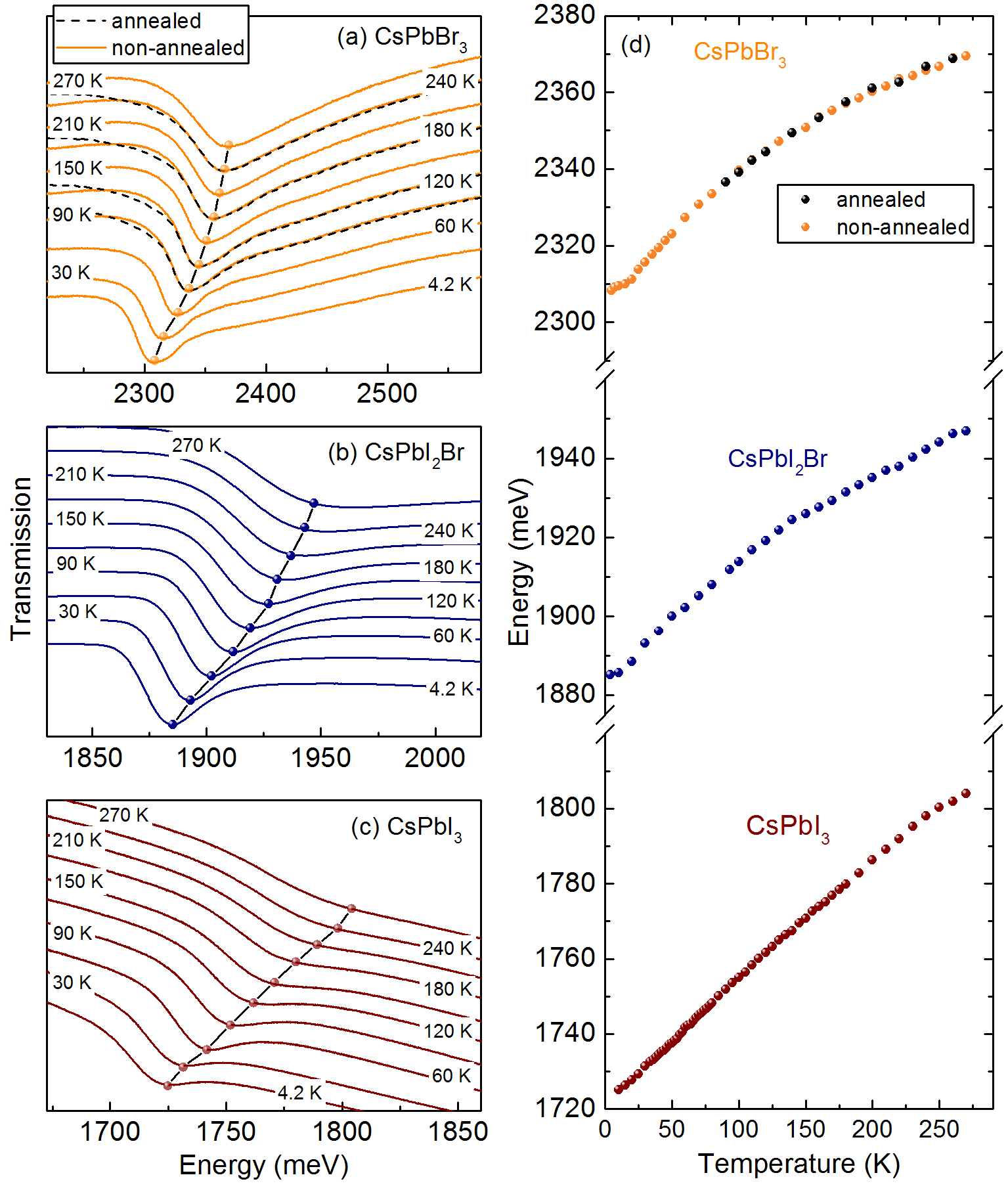}
\caption{Transmission spectra measured at different temperatures for
(a) CsPbBr$_3$, (b) CsPbI$_2$Br and (c) CsPbI$_3$. Symbols highlight
the evolution of the 1s absorption with temperature. (d) Energy of
the 1s transition (reflecting the evolution of the band gap) as a
function of temperature for the three compounds. For CsPbBr$_3$ data
is shown for the as prepared and annealed samples as described in
the text.} \label{fig:CST}
\end{figure}

Typical transmission spectra of CsPbBr$_3$, CsPbI$_2$Br and
CsPbI$_3$, measured over a wide range of temperatures (\SI{4.2}{\K}
- \SI{270}{\K}), are presented in Fig.\,\ref{fig:CST}(a-c). We
observe a consistent blue shift of the band edge absorption energy
of CsPbI$_3$ through mixed CsPbI$_2$Br to CsPbBr$_3$, highlighting
the good tuneability of the band gap via the introduction of a
heavier halide in the crystal \cite{sutton2016bandgap}. The
transmission spectrum for each compound exhibits a single minimum at
all temperatures, which blue shifts and broadens with increasing
temperature. The detailed evolution of the band gap absorption
energy with temperature is presented in Fig.\,\ref{fig:CST}(d). We
note that the band gaps of all the investigated samples exhibit a
well behaved monotonic dependence on the temperature. This is in
stark contrast with organic-inorganic halide perovskites, where an
increase in the band gap is observed at temperatures corresponding
to the phase transitions to a lower symmetry crystalline structure
\cite{Miyata15,Galkowski2016,yamada2015photoelectronic,d2014excitons,galkowski2017spatially,Yang2017}.
This is a particularly significant result in the case of CsPbI$_3$,
which suggests that our sample preparation procedure (see
Experimental Methods) preserves the photoactive perovskite phase of
this compound even at cryogenic temperatures. Importantly, we have
monitored the absorption spectra of all the samples during the cool
down. This point was critical especially for the CsPbI$_3$, which is
known to be unstable at ambient conditions. In the absorption
spectra we did not see any signs of dramatic change of the band gap,
which would suggest the transition into yellow phase. This supports
our finding that we freeze the sample in the cubic phase. In the
case of CsPbBr$_3$, we have investigated both as prepared samples
(orthorhombic phase) and samples, which have been annealed at
\SI{250}{\celsius}, which is above the transition to the cubic phase
\cite{Kulbak2015,stoumpos2013crystal}. The identical transmission
data and evolution of the gap with temperature shown in
Fig.\,\ref{fig:CST}(a,d) demonstrate that annealing does not
influence the electronic structure of CsPbBr$_3$ under the
measurements conditions used. This suggests that independently of
the thermal processing, CsPbBr$_3$ will always transform to the
orthorhombic phase below the phase transition point at
\SI{88}{\celsius} and the continuous evolution of its band gap with
temperature indicates the absence of any further phase transitions
and to a good stability of the investigated samples.

\begin{figure}[h]
  \centering
  \includegraphics[width= 1.0\linewidth]{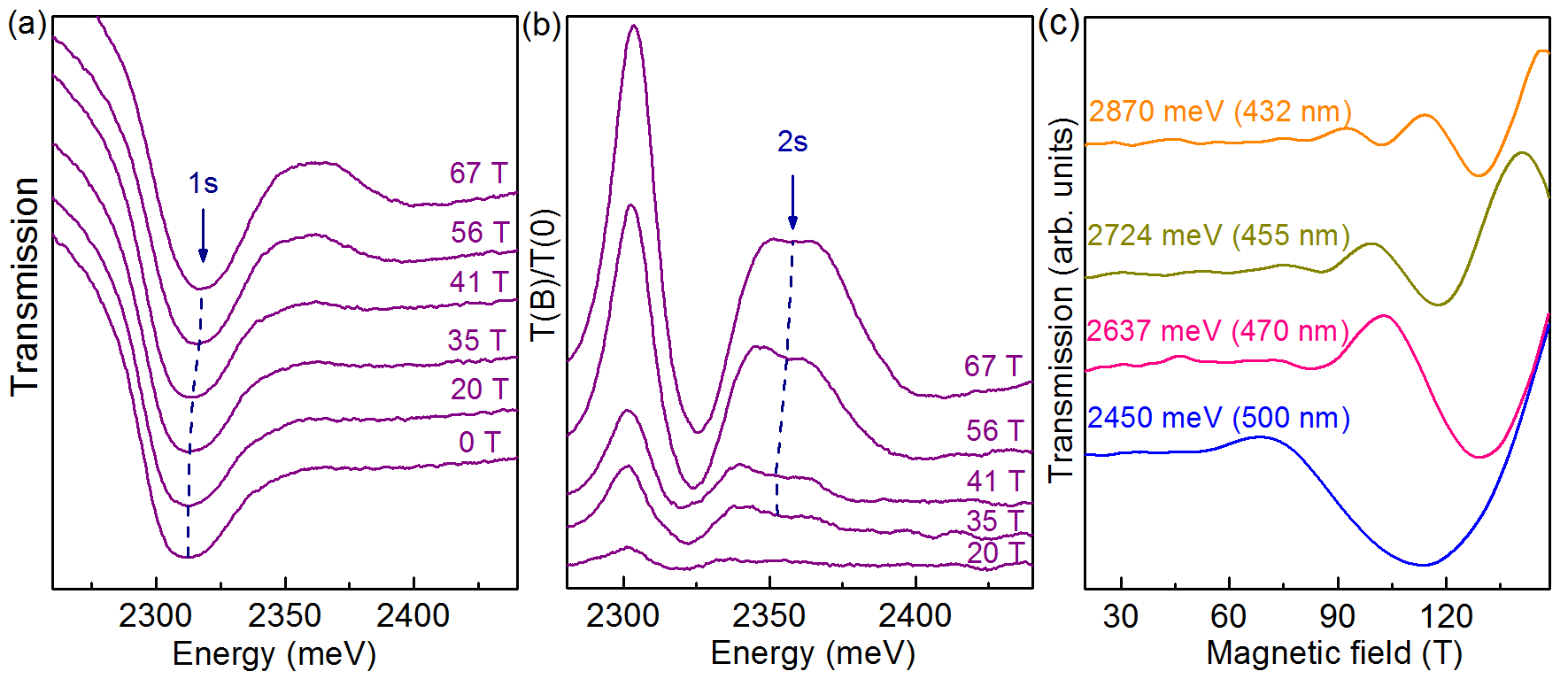}
     \caption{Low temperature transmission spectra in magnetic field for CsPbBr$_3$. (a) Transmission spectra measured at the indicated magnetic field values. (b) Transmission spectra in magnetic field divided by zero field spectrum. (c) Monochromatic transmission as a function of magnetic field obtained by the short pulse technique.}
  \label{fig:Br3SPE}
\end{figure}

Low temperature magneto transmission spectroscopy is a powerful
technique which has previously been used to precisely determine the
binding energy and reduced mass of the exciton in organic-inorganic
perovskites \cite{Miyata15,Galkowski2016,Yang2017}. In
Fig.\,\ref{fig:Br3SPE}(a) we present typical magneto transmission
spectra of CsPbBr$_3$ measured at \SI{2}{\K}. Related magneto
transmission spectra of CsPbI$_3$ and CsPbI$_2$Br are shown in the
Supporting Information (Fig1 SI). The pronounced minimum at
$\sim\SI{2313}{\milli\eV}$ is attributed to the 1s excitonic state.
With increasing magnetic field, we observe a clear blue shift of
this transition. On the high energy side of the 1s transition, a
weak minimum can be resolved for magnetic fields larger than
\SI{35}{\tesla}. It is more clearly seen in differential
transmission spectra obtained by dividing the transmission spectra
by the spectrum measured at zero magnetic field (see
Fig.\,\ref{fig:Br3SPE}(b)). At higher energy range, there are some
hints of absorption due to free carrier transitions between Landau
levels in the valence and conduction bands. To better resolve the
free carrier transitions, we have extended the long pulse
measurements to higher magnetic fields ($B>\SI{70}{\tesla}$) using
the short pulse generated by a single turn coil (see Experimental
Methods) to follow the absorption minima to higher energies.
Representative magneto transmission curves measured with a single
turn coil are presented in Fig.\,\ref{fig:Br3SPE}(c). These curves
show the evolution of the transmission through the sample of a
monochromatic (laser) light of different wavelengths measured as a
function of the magnetic field. The minima correspond to dipole
allowed transitions between Landau levels in the conduction and
valence band.

\begin{figure} [h]
\centering
\includegraphics[width= 1.0\linewidth]{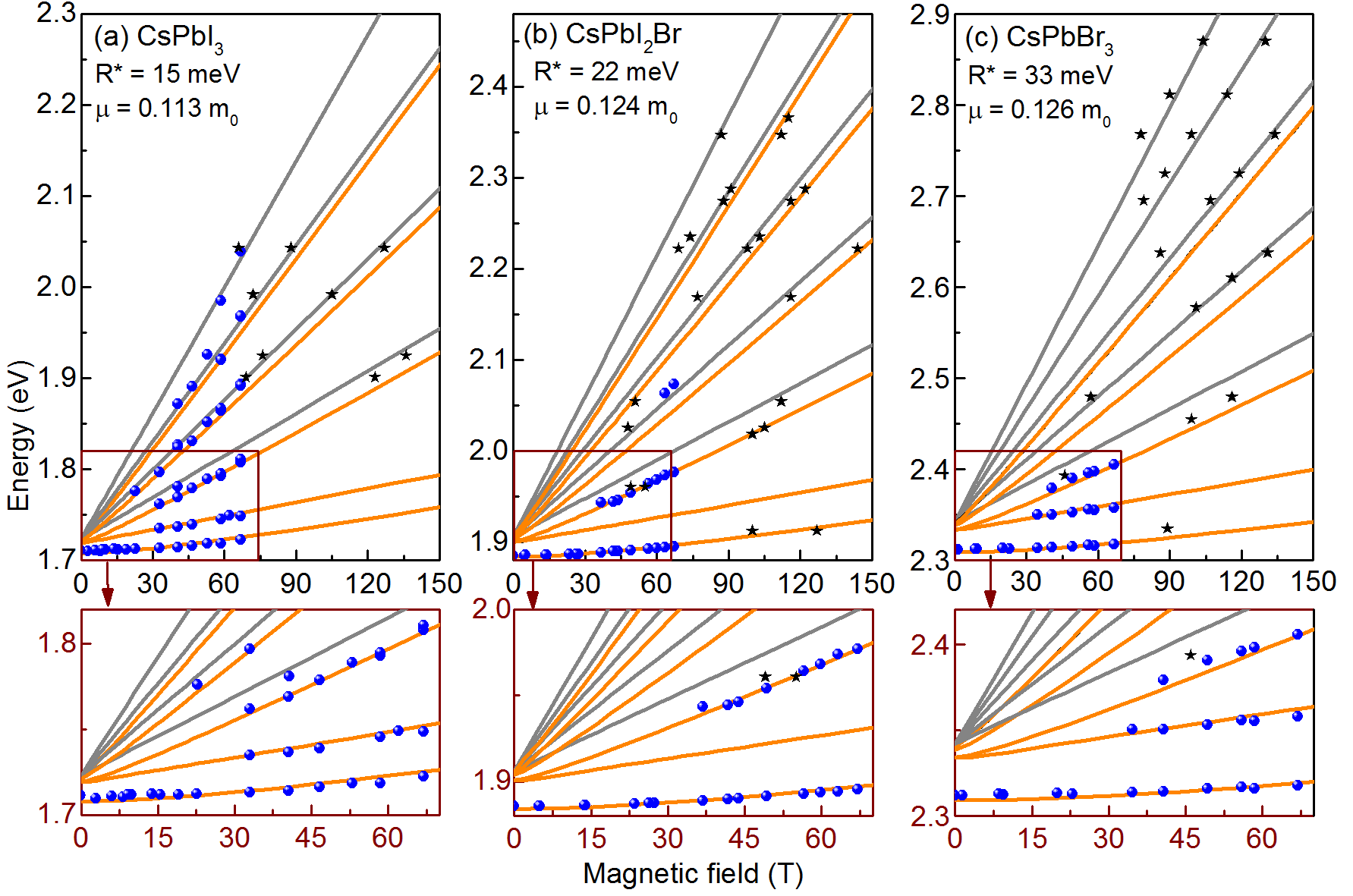}
\caption{Energy fan chart. Excitonic transition and inter Landau level transition energies as a function of magnetic field at
\SI{2}{\K} for (a) CsPbI$_3$, (b) CsPbI$_2$Br and (c) CsPbBr$_3$. Orange lines are the results of the fit to hydrogen-like
transitions. Grey lines indicate fitting result of the interband transition between Landau levels. Circles are data from long
pulse field measurements and stars are data from single-turn short pulse measurements. The lower panels show an expanded view of
the low field and low energy portion of the fan chart.}
  \label{fig:FC}
\end{figure}

To model the data further we plot the energetic positions of the
minima observed in the absorption spectra as a function of the
magnetic field, as marked by blue and black points in Fig.
\ref{fig:FC}. The analysis of the full magnetic field dependence of
both hydrogenic and free carrier transitions, shown in
Fig.\,\ref{fig:FC}, enables us to extract the exciton binding energy
and reduced mass \cite{Miyata15,Galkowski2016,Yang2017}. We
introduce a dimensionless parameter
$\gamma=\hbar\omega_{\text{c}}/2R^*$, where
$\omega_{\text{c}}=\text{e}B/\mu$ is the cyclotron frequency of
charge carriers and $\mu^{-1}=m_{\text{e}}^{-1}+m_{\text{h}}^{-1}$
defines the exciton reduced mass. In the high magnetic field limit
(corresponding to $\gamma>1$), the observed absorption resonances
are related to free carrier inter Landau level transitions with
energies
\begin{equation}
E = E_{\text{g}}+\left(n+\frac{1}{2}\right)\hbar\omega_{\text{c}}, \label{eq:LL}
\end{equation}
where $E_{\text{g}}$ is the band gap, $n=0,1,2,\dots$ is the orbital
quantum number of the Landau levels in the conduction and valence
bands. For dipole allowed transitions ($\Delta n=0$), and for a
well-defined value of the band gap $E_{\text{g}}$, the only fitting
parameter in Eq.(\,\ref{eq:LL}) is the reduced mass $\mu$. Fitting
the observed resonances to Eq.(\,\ref{eq:LL}) (gray lines in
Fig.\,\ref{fig:FC}) allows us to determine the exciton reduced mass.
The excitonic like transitions close to the band edge are well
described with a numerical model for a hydrogen atom in high
magnetic field \cite{Makado19}. The eigenenergies of an excitonic
system in zero field are given by
\begin{equation}
E_N = E_{\text{g}} - \frac{R^*}{N^2}, \label{eq:ExcitonEnergy}
\end{equation}
where $E_N$ is the energy of $N^{\text{th}}$ excitonic level, $R^* =
R_0{\mu}/m_0\varepsilon_{\text{eff}}^2$, $R_0$ is the atomic
Rydberg, $m_0$ is the free electron mass and
$\varepsilon_{\text{eff}}$ is the relative dielectric constant. The
fit of the inter Landau level transitions provides an accurate
estimation of $\mu$, which in this second step is taken as a fixed
parameter. This provides strong limits on the value of $R^*$, which
is further constrained by the observation of the 2s state, well
resolved only in high magnetic field both for CsPbI$_3$ and
CsPbBr$_3$. Contrary to early magneto optical estimates of exciton
binding energies of hybrid organic-inorganic perovskite
\cite{Hirasawa94,Tanaka03}, our approach does not require us to
assume a value for the effective dielectric constant, which we can
actually determine from Eq.(\ \ref{eq:ExcitonEnergy}). The values of
the effective mass and $\textit{R}^*$ obtained for all three
compounds are summarized in table\,\ref{Tab_Cs} and
Fig\,\ref{fig:GA}(a-b).

\begin{figure}[h!]
\centering
\includegraphics[width= 0.5\linewidth]{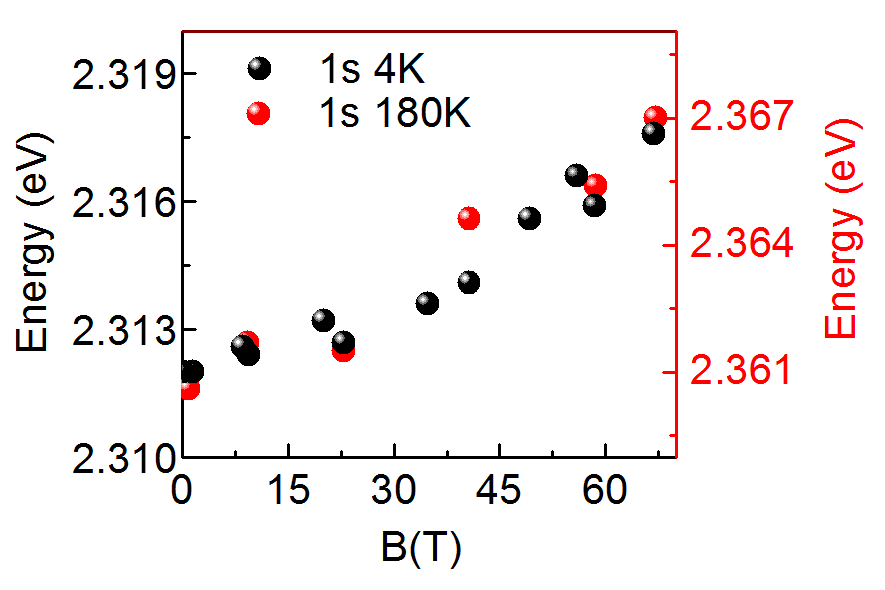}
\caption{Magnetic field dependence of 1s transition in CsPbBr$_3$ at two different temperatures.} \label{fig:comparisonLHT}
\end{figure}

\begin{figure}[h!]
  \centering
   \includegraphics[width= 0.5\linewidth]{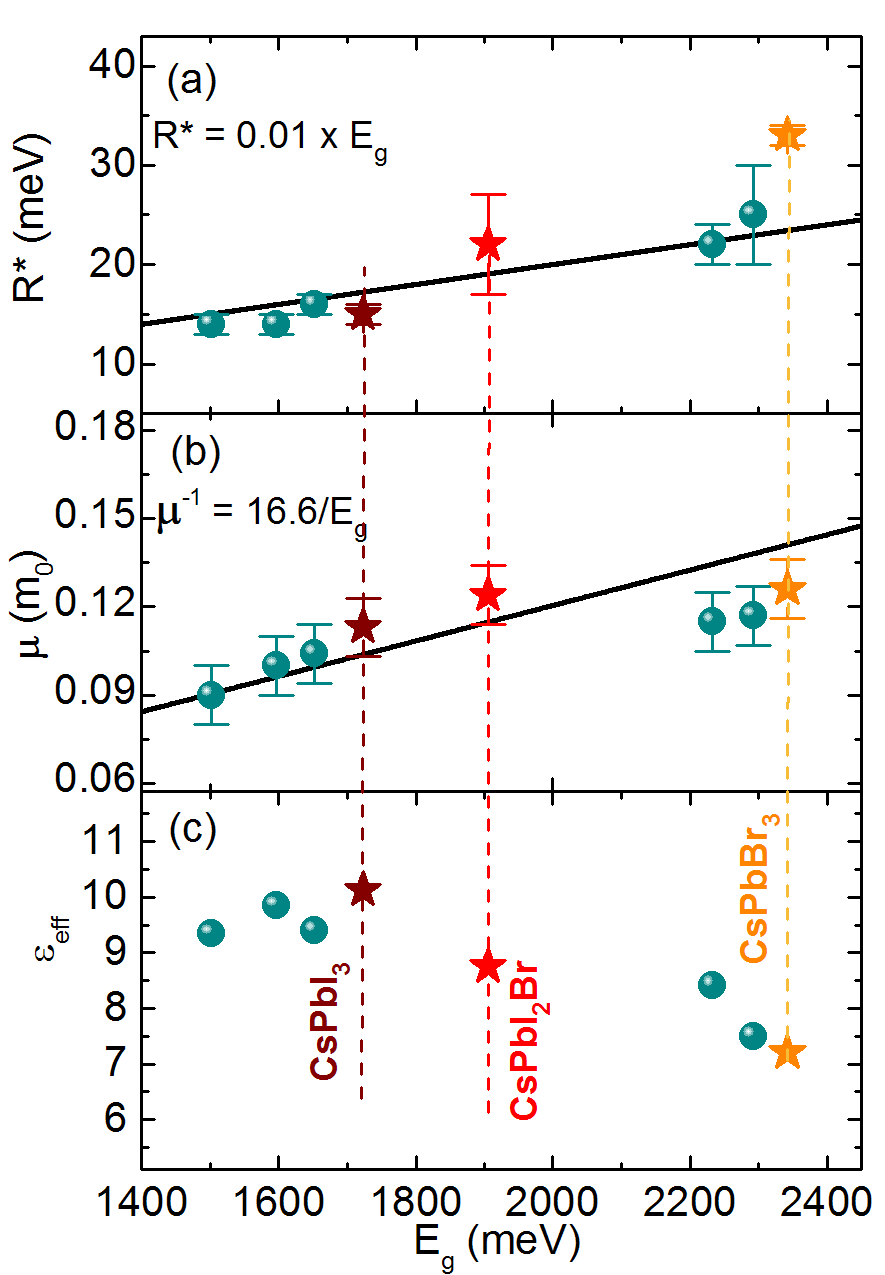}
\caption{(a) Binding energy, (b) effective mass and (c) dielectric
constant  as a function of the band gap. Brown, red and yellow stars
indicate the results for CsPbI$_3$, CsPbI$_2$Br and CsPbBr$_3$,
respectively. Blue symbols mark the results from previous work
\cite{Galkowski2016}.}
  \label{fig:GA}
\end{figure}

In the case of organic-inorganic perovskites, the phase transition
to a higher symmetry crystalline structure allows the rotational
motion of the organic cation, which enhances the dielectric
screening and reduces the exciton binding energy
\cite{Miyata15,Galkowski2016,Poglitsch87,Yang2017}. The lack of an
abrupt change of the band gap observed here implies that inorganic
perovskites do not undergo phase transitions up to room temperature.
The absence of a structural change suggests that the exciton binding
energy does not vary over the investigated temperature interval. To
support this conclusion, we have performed magneto-transmission
measurements of CsPbBr$_3$ at \SI{180}{\K}. We compare the magnetic
field dependence of the 1s transition energy at high and low
temperatures in Fig.\,\ref{fig:comparisonLHT}. The temperature
induced change in the band gap has been removed by plotting the high
temperature data on a different scale (right axis) but over the same
range (10\,meV in both cases). The excellent overlap of the two sets
of the data demonstrates that the exciton binding energy does not
change within experimental accuracy.

In Fig.\,\ref{fig:GA}(b) we plot the experimentally determined
values of the exciton reduced mass $\mu$ of Caesium compounds. Our
results are close to theoretical prediction of $\mu$ for CsPbI$_3$,
which range from $\sim0.07m_0$ \cite{chang2004first,amat2014cation}
to $\sim0.18m_0$ \cite{giorgi2014cation} and in excellent agreement
with density functional theory of organic-inorganic perovskites
adjusted to fit the experimental band gap \cite{Umari14}. We compare
our results on fully inorganic compounds to the reduced masses
determined on organic-inorganic perovskites
\cite{Miyata15,Galkowski2016,Yang2017}. We observe an increase of
the effective mass with increasing value of the band gap. This trend
can be understood in the frame of a simple two-band
$\mathbf{k}\cdot\mathbf{p}$ model \cite{even2015solid}, assuming the
same effective mass for the valence and conduction band, the reduced
mass of the exciton can be written as
\begin{equation}
\frac{1}{\mu}=\frac{4 | P|^2}{m_0 E_{\text{g}}},
\label{eq:ReducedMass}
\end{equation}
where $P=\langle \Psi_{\text{VB}}| p_x | \Psi_{\text{CB}}\rangle$ is
the momentum matrix element which couples states in the conduction
and valence bands and $2| P|^2/m_0$ is the Kane energy
\cite{Galkowski2016}. The measured evolution of the band gap with
the reduced mass is well fitted with a Kane energy of \SI{8.3}{\eV},
only slightly larger than theoretical predictions \cite{Even14}.

Knowing the values of exciton binding energy and the reduced
effective mass we can calculate the effective dielectric constant
$\varepsilon_{\text{eff}}$. In Fig.\,\ref{fig:GA}(c), we compare the
dielectric constant for Caesium-based compounds (stars) with the
corresponding quantities for the hybrid organic-inorganic materials
from our previous work \cite{Miyata15,Galkowski2016,Yang2017} (blue
circles). The dielectric constant does not vary significantly for a
given lead-halide cage, regardless of the cation, while it decreases
with decreasing halide mass. This is consistent with theoretical
calculations of the complex dielectric function of MA- and CsPbI$_3$
\cite{berdiyorov2016role} and with the computed dielectric functions
of MASnX$_3$ \cite{feng2014effective} and MAPbX$_3$
\cite{feng2014crystal}, which suggest that the main contributions to
the dielectric permittivity are due to Pb-X stretching modes and
Pb-X-Pb rocking modes \cite{perez2015vibrational}. Detailed Raman
spectroscopy on the MAPbX$_3$ series (with X=I, Br, Cl) has
demonstrated that both Pb-X stretching and Pb-X-Pb rocking modes
harden with decreasing halide mass, with the derived dielectric
function exhibiting the same trend as observed in
Fig.\,\ref{fig:GA}(a) \cite{sendner2016optical}. This confirms the
qualitative picture that the dielectric screening properties are
mainly determined by the lead-halide cage in the low temperature,
when the organic cation motions are frozen.

The exciton binding energies we have determined are significantly
smaller than early estimates obtained from magneto optical
measurements of the 1s excitonic states \cite{Hirasawa94}. In the
absence of any observed phase transitions (see Fig.\ \ref{fig:CST}),
we argue that the exciton binding energy will depend only very
weakly on the temperature, as demonstrated by the similar
diamagnetic shift of the 1s state at \SI{2}{\K} and \SI{170}{\K} in
Fig.\ \ref{fig:comparisonLHT}. This suggests that the photo-created
carriers in the compounds investigated exhibit essentially a
free-carrier behavior at temperatures corresponding to the normal
operating conditions of solar cells.

\begin{table}[h]
\begin{center}
\caption{Parameters determined from the fit of the full Landau fan
chart for Caesium-based compounds at $\SI{2}{\K}$.}\label{Tab_Cs}
\begin{tabular}{lccccc}
\hline
Compound & Phase & $E_{\text{g}}$ (meV) & $R^{*}$ (meV)& $\mu$  $(\text{m}_0)$ & $\varepsilon_{\text{eff}}$ \\
\hline
CsPbI$_{3}$ & cubic & 1723 & $15\pm1$ & $0.114\pm0.1$ & 10.0 \\
CsPbI$_{2}$Br & cubic & 1906 & $22\pm3$ & $0.124\pm0.02$ & 8.6 \\
CsPbBr$_{3}$ & orthorhombic & 2342 & $33\pm1$ & $0.126\pm0.01$ & 7.3 \\
\hline
\end{tabular}
\end{center}

\end{table}

In summary, we have presented a detailed magneto-optical investigation of fully inorganic CsPbX$_3$ perovskites (X=I, Br and a
mixture of these). The well behaved (continuous) temperature dependence of the band gap points to the absence of any structural
phase transitions for perovskites with fully inorganic cations. By performing magneto transmission spectroscopy up to
\SI{150}{\tesla}, we have determined key electronic parameters including the exciton binding energy and reduced mass with high
accuracy. Our approach does not require any assumption concerning the strength of the dielectric screening, which we determine a
posteriori. The comparison of the values of the dielectric constant for inorganic and hybrid perovskites suggests that the dominant
contribution to dielectric screening is related to the relative motion within the lead-halide cage. For the fully inorganic
compounds, the values of $R^*$ and $\mu$ increase with the band gap energy, similarly to hybrid perovskites. We conclude that at
low temperature, when the organic cations are ordered, the qualitative picture of the interactions within the lattice is
essentially the same for both the inorganic and hybrid compounds. This is consistent with the ability to optimize perovskite solar cells by adjusting the cation structure while maintaining the same device operation modes.

\section{Experimental Methods}
Magneto transmission spectra have been acquired by combining long
pulse magnetic field measurements for magnetic fields up to
\SI{66}{\tesla} and short duration pulsed magnets (for magnetic
fields up to \SI{150}{\tesla}). For the long pulse measurements
(typical pulse duration $\sim\SI{100}{\milli\s}$), the sample was
mounted in a liquid helium cryostat. White light from a halogen lamp
was used as the excitation source. The light emitted from the lamp
was coupled in a \SI{200}{\micro\m} diameter multimode fiber, used
to illuminate the sample. The transmitted light was coupled in a
\SI{400}{\micro\m} diameter multimode fiber and guided to
spectrometer equipped with a liquid nitrogen cooled CCD camera. The
typical exposure time was 2ms, which ensured that the transmission
spectra were acquired at essentially constant magnetic field values.
For the very high magnetic field measurements
($B<\SI{150}{\tesla}$), magnetic field pulses with a typical
duration of \SI{5}{\micro\s} were generated by a single turn coil
system with a bore diameter of \SI{10}{\milli\m}. A helium-flow
cryostat with a kapton tail was located in the single turn coil. The
sample was kept at \SI{5}{\K}. Magneto-transmission measurements
were conducted by using a tunable optical parametric oscillator
(OPO) pumped by a Ti:sapphire laser as the light source, a fast
(\SI{100}{\MHz}) silicon detector and a high speed digital
oscilloscope.

All samples were prepared on glass microscope slides, which were
cleaned by sonication sequentially in acetone and isopropanol, and
then treated with oxygen plasma for 10 minutes.  The CsPbI$_{3}$ and
CsPbI$_{2}$Br perovskite layers were deposited in a nitrogen
glovebox by spincoating (at 1500 rpm) a solution of the appropriate
ratios of CsI (Alfa Aesar, 99.9$\%$), PbI$_{2}$ (Sigma Aldrich,
99$\%$) and PbBr$_{2}$ (Sigma Aldrich, $>$98$\%$) dissolved at 0.43
M in N,N-Dimethylformamide following a previously reported procedure
\cite{sutton2016bandgap}. The CsPbBr$_{3}$ perovskite was deposited
by sequential evaporation of layers of PbBr$_{2}$ (107 nm) and CsBr
(93 nm), each deposited at 1-2 ${\AA}$/s onto room-temperature
substrates, at pressures below $6 \times 10^{-6}$ mbar in a BOC
Edwards Auto 306 evaporator. Owing to their limited stability in air
at room temperature, CsPbX$_3$ samples were systematically annealed
in an oven before loading them in the cryostat in order to restore
the cubic phase. CsPbI$_3$ was annealed at \SI{350}{\celsius} for 10
minutes. CsPbI$_2$Br and CsPbBr$_3$ were annealed at
\SI{250}{\celsius} for 5--7 minutes. After the annealing, the
samples were placed in a liquid helium cryostat within 4 minutes,
which ensured that the sample remains in the cubic phase. The sample
stage was not temperature-controlled.  All samples were annealed
after deposition and again before measurement.

\begin{acknowledgement}

This work was partially supported by ANR JCJC project milliPICS, the
R{\'e}gion Midi-Pyr{\'e}n{\'e}es under contract MESR 13053031,
BLAPHENE and TERASPEC project which received funding from the IDEX
Toulouse, Emergence program,  NEXT ANR-10-LABX-0037 in the framework
of the ``Programme des Investissements d'Avenir''. Z. Y. held a
fellowship from the Chinese Scholarship Council (CSC), R. J. S. is a
Commonwealth Scholar, funded by the UK government. This work was
supported by EPSRC (UK) via its membership to the EMFL (grant no.\
EP/N01085X/1).

\end{acknowledgement}

\begin{suppinfo}
Supporting information: Low temperature transmission spectra in
magnetic field for CsPbI$_{2}$Br and CsPbI$_{3}$, XRD, Absorbance
and PL for CsPbI$_{3}$ CsPbI$_{2}$Br and CsPbBr$_{3}$.

\end{suppinfo}

\bibliography{CsPerovskite}


\end{document}